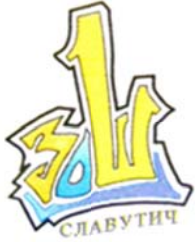

Юля Нога, Віка Лутченко, Настя Грань, Соня Миткевич
6 клас, школа №1, місто Славутич, Київська область
**керівник – Володимир Шаталов**

# ШКОЛА МАЙБУТНЬОГО

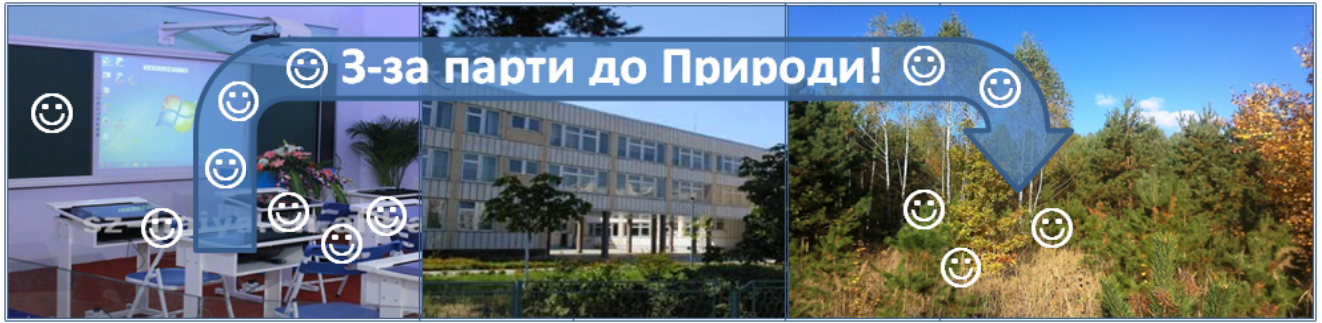

*Навчатися у самої Природи*

## Нові поступки

### Розширення занять поза межі класної кімнати

Все теперішнє покоління учнів мріє, щоб школи працювали на електроніці. Вже зараз на уроках у деяких школах використовують комп'ютери, мультімедійни проектори, електронні сенсорні дошки та інше. Як нам здається, в школі майбутнього застосування інформаційних технологій може бути розширене за межі занять у класі з тім, щоб навчатися у самої Природи. Вже зараз для цього існують усі необхідні пристрої та програмне забезпечення.

Майже кожний школяр має у себе в кишені смартфон, як наприклад, Samsung Galaxy. Насправді, цей пристрій може бути перетворений за допомогою спеціальних програм у маленьку кишенькову лабораторію, що дозволяє учням виконувати лабораторні роботи і наукові спостереження. Наприклад, за допомогою звичайного смартфона на якому встановленні спеціальні програми можна вимірювати такі фізичні параметри навколишнього середовища, як радіаційний фон [1], рівень електромагнітного забруднення [2], рівень освітлення, шуму [3], та інше. Причому, усі ці параметри за допомогою системи глобального позиціонування (GPS) прив'язані до місця спостереження. Водночас, ці смартфони можуть бути застосовані для вимірювання деяких показників здоров'я людини, наприклад, серцевий ритм [4] і кров'яний тиск [5]. Прикладні програми, що необхідні для цих спостережень, є у вільному доступі у службі Google Play.

### Спільне навчання поза межами школи, міста, країні

Друга можливість, яка надається учням школи майбутнього за допомогою існуючих інформаційних технологій, це безперервне і дистанційне навчання за допомогою Інтернету, яке не обмежується часовими рамками шкільного розкладу уроків і просторовими межами класної кімнати. Групи учнів також можуть бути розширені за межі школи, міста чи країни.

Використання Інтернету дає можливість проводити уроки на природі або в музеї. При цьому учні за допомогою смартфона створюють аудіо чи відео звіти та відправляють їх на перевірку.

# Автоматичне спостереження за станом здоров'я та навчальною активністю учнів

Налаштовані спеціальним чином смартфони дають можливість відслідковувати на відстані стан здоров'я школярів при виконанні ними самостійної роботи або на прогулянках, в походах, під час тренувань. Ці дані, як і результати роботи або тренувань, за допомогою Інтернету передаються на шкільний комп'ютер, який здійснює їх обробку і, в разі будь-яких відхилень, посилає учневі поради або дає сигнал тривоги, повідомляє вчителю або батькам.

## Як це працює

### Смартфон

Більшість смартфонів (але не всі) оснащені різноманітними датчиками, а саме – датчиками тиску, вологості, освітленості, лінійного прискорення, температури, наближення, сили тяжіння, а також мають акселерометр, гіроскоп, кутомір, барометр, магнітометр та інші мікро-електро-механічні системи (MEMS). Перевірити, які датчики знаходяться у смартфоні, можна за допомогою програми Sensor Kinetics [6]. Ця програма виконує також і освітні функції, вона демонструє фізику гравітації, що таке прискорення, обертання, магнетизм та інші сили, які вимірюються за допомогою смартфона. За завданням учителя школярі можуть проводити різні експерименти, які виконуються і перевіряються за допомогою датчиків. Інтерпретація показань цих датчиків здійснюється за допомогою прикладних програм на основі фізичних законів, якими описуються спостережувані явища. Робота з цими програмами змушує учнів вивчати ці закони, що допомагає закріпленню шкільного матеріалу на практиці.

### Інтернет

Отримані на смартфонах різних учнів дані за допомогою Інтернету передаються в «хмарну» інфраструктуру, наприклад, Google Disk, яка призначена для вільного використання широким загалом. Там вони обробляються і об'єднуються в єдину базу даних, після чого їх може перевірити вчитель. Крім того, до цих даних отримують доступ всі учасники. Ці дані можна наносити на мапи для створення картини екологічного стану як прилеглих до школи територій, так і інших регіонів, що беруть участь у проекті. Проведені таким чином вимірювання та експерименти можуть послужити основою для організації спільних проектів шкільних і наукових досліджень, що не обмежуються одною школою, містом, або країною.

### Зворотній зв'язок

Отримані із «хмари» дані можна обробляти в одному, або декількох центрах. Це може бути шкільний комп'ютер, а також будь-який інший. Ці центри аналізують результати вимірів і видають рекомендації учням, допомагають вчителю оцінити їх роботу, батькам простежити за самостійною роботою дітей, а шкільним лікарям допоможуть вчасно виявити у дітей можливі проблеми зі здоров'ям.

### При вивченні яких предметів це можна використовувати

На нашу думку, вимірювання за допомогою смартфонів можуть покращити процеси вивчення таких предметів.

*Математика*

Знайомство з законами геометрії, тригонометрії та стереометрії на місцевості за допомогою вимірювача відстаней і кутів.

*Фізика*

Наочне спостереження деяких явищ, що описуються законами фізики та експериментальна перевірка відповідних математичних формул. Спостереження і вимір невидимих випромінювань і електромагнітних сил.

*Природознавство*

Наочне спостереження, фото і відео фіксація змін навколишнього середовища під впливом життєдіяльності людей.

*Географія*

Практичне освоєння геоінформаційних технологій шляхом нанесення на карту місцевості результатів вимірювань параметрів навколишнього середовища.

*Біологія*

Знайомство і спостереження за серцевими ритмами і процесами, які регулюють кровообіг в організмі людини.

*Фізкультура*

Облік ефективності тренувань, стеження за пульсом і кров'яним тиском залежно від навантаження.

### Як це допоможе школі

Процес навчання частково перетворюється у гру та, водночас, створює інформаційну базу, яку можна використовувати де завгодно - і в школі, і в місцевому самоврядуванні, і в екологічному моніторингу, і в наукових дослідженнях, які пов'язані з моніторингом глобальних змін параметрів навколишнього середовища.

## Висновки

Запропоновані нові можливості використання інформаційних технологій у навчальному процесі школи майбутнього, а саме, розширення занять поза межі класної кімнати; спільне навчання і дослідження поза межами школи, міста, країні; автоматичне спостереження за станом здоров'я та навчальною активністю учнів.

Усі ці ідеї придатні до втілення в учбовий процес вже зараз, цілком або частково, що демонструють наші попередні випробування.

Ці новації допомагатимуть вирішувати проблеми забезпеченості шкіл комп'ютерною технікою, оскільки кожний школяр використовує свій смартфон. Також дані, що отримані за допомогою цих пристроїв та поширені за допомогою мережі Інтернет, можуть статися в нагоді при вирішенні екологічних проблем регіону, та навіть країни.

Запропонована технологія участі школярів у зборі локальних параметрів екологічного стану різних міст здатна суттєво вплинути на творчу активність школярів, що підвищує рівень навчального процесу. Проведення спільних вимірювань учнями різних класів або шкіл створює дух змагань, що завжди стимулює навчальний процес.

Таким чином, відмітною особливістю школи майбутнього буде використання інформаційних технологій у навчальному процесі з метою покращення рівня навчання, а також доступ до сучасних технологій більшій кількості учнів. За допомогою новітніх технологій і техніки можна змінити життя на краще, зробити навчальний процес цікавішим та продуктивнішим.

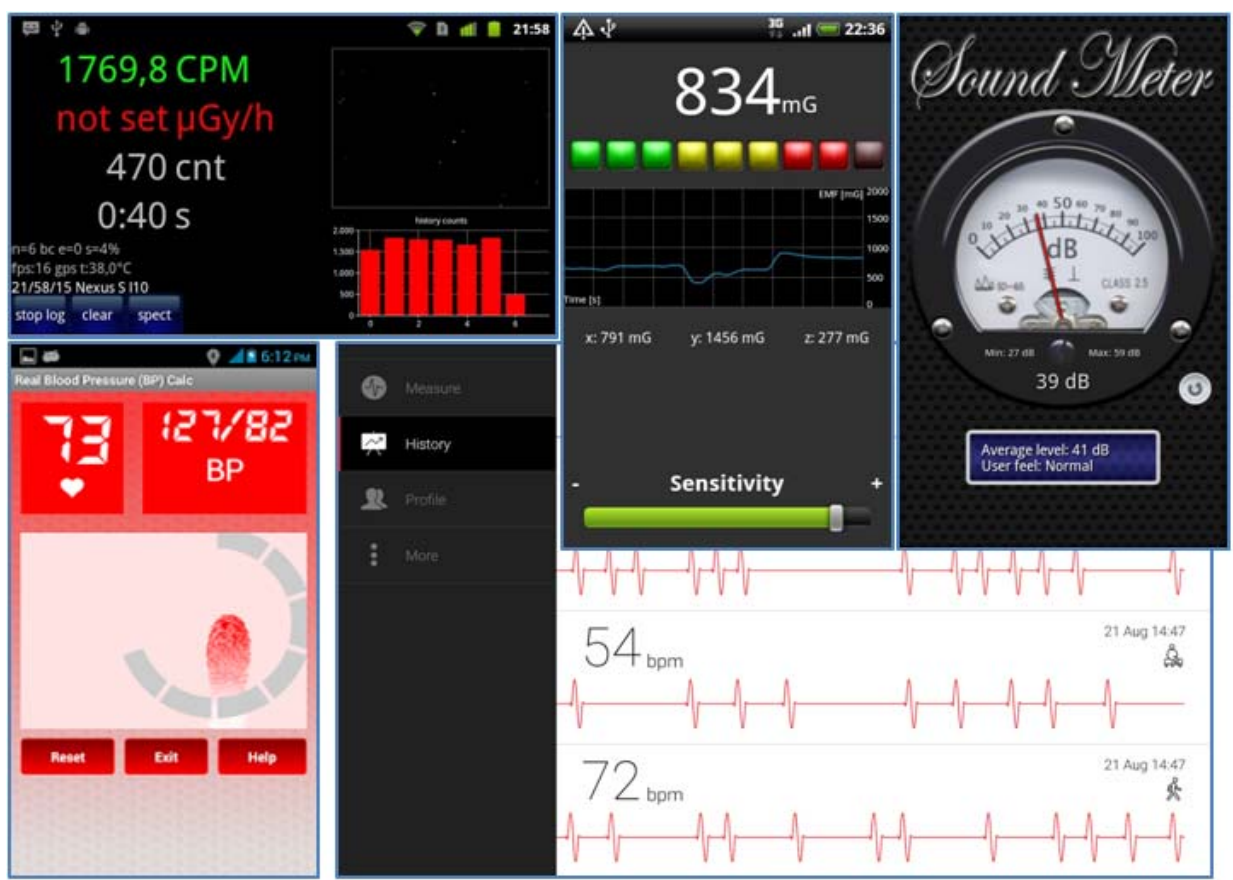

*СМАРТФОН ЯК КИШЕНЬКОВА ЛАБОРАТОРІЯ*

# Посилання